\documentclass[11pt,twoside]{article}
\usepackage{asp2014}
\usepackage{enumitem}
\setlist{noitemsep}
\aspSuppressVolSlug
\resetcounters

\begin{document}
\title{Annotating TAP responses on-the-fly against an IVOA data model}
\author{Mireille~Louys,$^{1,2}$ Laurent~Michel,$^2$ Fran\c cois~Bonnarel,$^2$ and Joann~Vetter$^3$ }

\affil{$^1$Universit\'e de Strasbourg, ICube, CNRS UMR 7357, Strasbourg, France \email{mireille.louys@unistra.fr}}
\affil{$^2$Universit\'e de Strasbourg, Observatoire Astronomique de Strasbourg, CNRS UMR 7550, Strasbourg, France}
\affil{$^3$Universit\'e d' Evry, ENSIIE, EVRY, France}

\paperauthor{Louys Mireille}{mireille.louys@unistra.fr}{0000-0002-4334-1142}{University of Strasbourg}{ICube UMR7357-CNRS, Observatoire Astronomique UMR7357-CNRS} {Strasbourg}{}{67000}{France}
\paperauthor{Bonnarel Fran\c{c}ois}{francois.bonnarel@astro.unistra.fr}{}{University of Strasbourg}{Observatoire Astronomique,  UMR7550-CNRS}{Strasbourg}{}{67000}{France}
\paperauthor{Laurent Michel}{laurent.michel@astro.unistra.fr}{0000-0001-5702-0019}{University of Strasbourg}{Observatoire Astronomique UMR7550-CNRS} {Strasbourg}{}{67000}{France}
\paperauthor{Joann Vetter}{joann.vetter@ensiie.fr}{}{University d'Evry}{ENSIIE}{EVRY}{}{91025}{France}



\begin{abstract} 

With the success and widespread of the IVOA Table Access Protocol (1) for discovering and querying tabular data in astronomy, more than one hundred of TAP services exposing altogether 22 thousands of tables are accessible from the IVOA Registries at the time of writing. Currently the TAP protocol presents table data and metadata via a {TAP\_SCHEMA} describing the served tables with their columns and possible joins between them. We explore here how to add an information layer, so that values within table columns can be gathered and used to populate instances of objects defined in a selected IVOA data model like Photometry, Coords, Measure, Transform or the proposed MANGO container model. 
This information layer is provided through annotation tags which tell how the columns' values can be interpreted as attributes of instances of that model. Then when a TAP query is processed, our server add-on interprets the ADQL query string and produces on-the-fly, when possible, the TAP response as an annotated VOTable document. The FIELD elements in the table response are mapped to corresponding model elements templated for this service. This has been prototyped in Java, using the VOLLT package library and a template annotation document representing elements from the MANGO data model. This has been exercised on examples based on Vizier and Chandra catalogs.

\end{abstract}

\section{Goal }

Today a large collection of services distributing table data via the TAP access protocol \citep{2019ivoa.spec.0927D} are provided within the IVOA framework. 
In this paper we focus on source catalogs, served by TAP. 
Catalogs are enriched nowadays with associated data added to the astronomical source's measurements, like previews or images from other collections on which it can be located, a spectrum or a time series recorded for this object, etc. 
TAP relies on the standardized VOTable format (\citep{2019ivoa.spec.1021O} to present TAP responses but it uses free names for table columns, so these are not homogeneous across various TAP services. 
However data interpretation relies on general quantities like coordinates in spatial and spectral domains, photometric flux or densities,velocity, time stamps for events, etc. 
Those concepts are represented in IVOA data models as Coords \citep{2021ivoa.spec.Coords}, Measurement \citep{2021ivoa.spec.Meas}, Photometry \citep{2013ivoa.spec.1005S}. 
The data associated to a source, and linked to a catalog entry can also be described using 'data products-oriented' data models like ObsCore 
, Spectrum or Cube data models.
In a new attempt to unify the source measurements together with associated data sets in a common Object Oriented interface, we have designed MANGO \citep{MANGO} to describe the various facets contained in a source catalog.


This is very useful when dealing with enriched catalogs.
Data models carry an elaborated view on catalog data and help to represent and trace various kinds of data and metadata such as : 
\begin{itemize}
\item catalog data with groups of columns interpreted as objects 
\item detailed calibration metadata in terms of astrometry, photometry, spectral and temporal calibration, etc.
\item classification or quality flags
\item data products attached to a detection (source) and distributed by another service like spectrum, spectral energy distribution, light-curve, image, cube, etc.
\end{itemize}
\flushleft

We exercised an annotation framework using:
\begin{itemize}
 \item the ModelInstanceinVOTable specification syntax (in development see \citep{github:mapping}) to represent annotated objects from MANGO
 \item a JSON template format to express how TAP columns can be encapsulated as attributes of objects in the data model logic
\end{itemize}

This has been implemented on a prototype TAP service with the CDS-TAP library \citep{CDSTAPLib}.
The VOLLT TAP toolkit \citep{Vollt}  allows to customize the "writeHeader" method in order to decorate the VOTable TAP response with a MANGO annotation block in XML, in compliance to the JSON template provided.
From the VODML-XML representation of each data model, a list of model elements can be extracted to form the list of XML components for annotating TAP outputs for one IVOA Model. A JSON dictionary of components, can be derived and adjusted by the data provider to each table served with additional references to the FIELDS provided in the TAP response. 

Fig.\ref{fig:annot} summarizes the usual TAP query mechanism (in green) and highlights the role of the developed Mapping Engine which incorporates in the annotation the XML components snippets based on data model elements and mentioned as used elements in the provider's JSON profile.

\articlefigure[width=\textwidth]{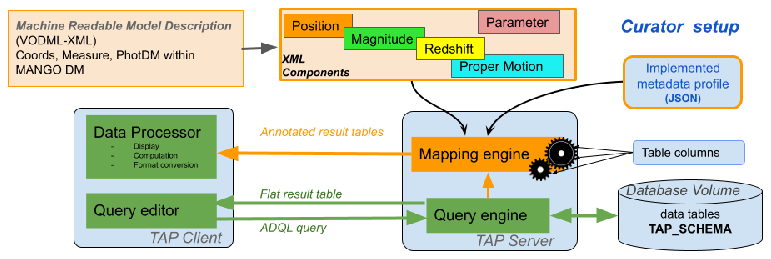}{fig:annot}{Annotation process on top of the TAP query scenario : in green the usual query scenario, in orange, the elements added for the annotation.}

\section{Scenario description}

Fig.\ref{fig:workflow} illustrates the various steps involved for creating the list of XML components,
analysing the selection in the query and building the annotation block.

\articlefigure[width=\textwidth]{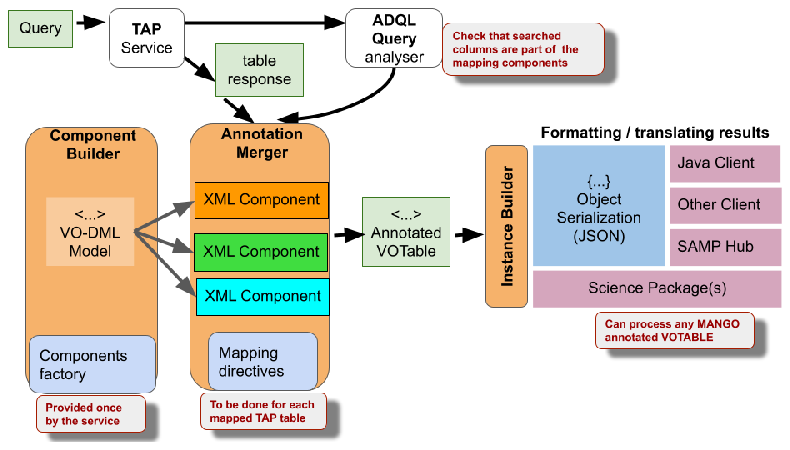}{fig:workflow}{The building blocks used for the mapping generation. From left to right: the Component Builder, the Annotation Merger and the Instance Builder. 
\footnotesize {Cf. presentation by L.Michel, IVOA meeting May 2020, A Component and Association Based Model For Source Data}}

Here is the summary of a typical annotation scenario to run on the prototype: 
\begin{itemize}
    \item  A TAP query is processed by the server which prepares results in a VOTable TABLE structure.
    \item The Annoter gets the appropriate annotation profile (currently in JSON format) defined by the data curator as a list of components served by the TAP server. It contains the binding of data model elements and column references to reach data model leaves.
    \item The TreeWalker program browses this annotation profile. It compares the FIELDs elements in the VOTable response with their counterparts attributes in the model elements represented. Relevant XML components are identified and appended into the annotation tree.
    \item The Annoter wraps this annotation block as a VOTable resource and inserts it at the top of the usual VOTable TAP response. 
\end{itemize}
Examples of the various files run with the prototype are available at \url{https://github.com/loumir/TAP-annoter/AdassProceedingsX3-010/}. 
\section {Parsing the annotated tables}

After an annotation is produced and inserted on top of the VOTable response, a client application can reuse it in different ways. For instance for checking errors only, or for building full objects instances.

\section{Conclusion}
The prototype serves as a proof of concept for the wrapping of TAP responses with IVOA models' metadata, here MANGO and the {Coords}, {Meas}, {Photometry} data models.
The format used to describe the library of mapping components can be either XML or JSON. 
We foresee development for an integration of such a strategy in the PyVO framework.

\acknowledgements For the support of ESCAPE (European Science Cluster of Astronomy and Particle Physics ESFRI Research Infrastructures) funded by the EU Horizon 2020 research and innovation program (Grant Agreement n.824064).
\bibliography{X3-010}  

\begin{thebibliography}{}
\expandafter\ifx\csname natexlab\endcsname\relax\def\natexlab#1{#1}\fi
\expandafter\ifx\csname url\endcsname\relax
  \def\url#1{\texttt{#1}}\fi
\expandafter\ifx\csname urlprefix\endcsname\relax\def\urlprefix{URL }\fi
\providecommand{\eprint}[2][]{\url{#2}}

\bibitem[{{Cresitello-Dittmar} \& {Rots}(2021)}]{2021ivoa.spec.Meas}
{Cresitello-Dittmar}, M., \& {Rots}, A. 2021, {Astronomical Measurements Model,
  Version 1.0}, IVOA Recommendation to appear.
  \urlprefix\url{https://www.ivoa.net/documents/Meas/}

\bibitem[{{Dowler} et~al.(2019){Dowler}, {Rixon}, {Tody}, \&
  {Demleitner}}]{2019ivoa.spec.0927D}
{Dowler}, P., {Rixon}, G., {Tody}, D., \& {Demleitner}, M. 2019, {Table Access
  Protocol Version 1.1}, IVOA Recommendation 27 September 2019

\bibitem[{{Mantelet}(2019)}]{CDSTAPLib}
{Mantelet}, G. 2019, {TAP Library v2.3 }, CDS portal software for IVOA.
  \urlprefix\url{http://cdsportal.u-strasbg.fr/taptuto/}

\bibitem[{{Mantelet}(2021)}]{Vollt}
--- 2021, {VOLLT software for TAP }, CDS portal software for IVOA.
  \urlprefix\url{https://github.com/gmantele/vollt}

\bibitem[{{Michel} et~al.(2021{\natexlab{a}}){Michel}, {Cresitello-Dittmar},
  Bonnarel, Landais, Louys, Salgado, \& Lemson}]{github:mapping}
{Michel}, L., {Cresitello-Dittmar}, M., Bonnarel, F., Landais, G., Louys, M.,
  Salgado, J., \& Lemson, G. 2021{\natexlab{a}}, Model instances in votables,
  \url{https://github.com/ivoa-std/ModelInstanceInVot}

\bibitem[{{Michel} et~al.(2021{\natexlab{b}}){Michel}, Landais, Bonnarel,
  Louys, Salgado, \& Molinaro}]{MANGO}
{Michel}, L., Landais, G., Bonnarel, F., Louys, M., Salgado, J., \& Molinaro,
  M. 2021{\natexlab{b}}, {MANGO: A Component and Association Based Model for
  representing data for astronomical sources, Version 1.0}, in development.
  \urlprefix\url{https://github.com/ivoa-std/MANGO}

\bibitem[{{Ochsenbein} et~al.(2019){Ochsenbein}, {Taylor}, {Donaldson},
  {Williams}, {Davenhall}, {Demleitner}, {Durand}, {Fernique}, {Giaretta},
  {Hanisch}, {McGlynn}, {Szalay}, \& {Wicenec}}]{2019ivoa.spec.1021O}
{Ochsenbein}, F., {Taylor}, M., {Donaldson}, T., {Williams}, R., {Davenhall},
  C., {Demleitner}, M., {Durand}, D., {Fernique}, P., {Giaretta}, D.,
  {Hanisch}, R., {McGlynn}, T., {Szalay}, A., \& {Wicenec}, A. 2019, {VOTable
  Format Definition Version 1.4}, IVOA Recommendation 21 October 2019

\bibitem[{{Rots} et~al.(2021){Rots}, {Cresitello-Dittmar}, \&
  {Laurino}}]{2021ivoa.spec.Coords}
{Rots}, A., {Cresitello-Dittmar}, M., \& {Laurino}, O. 2021, {Astronomical
  Coordinates and Coordinate Systems, Version 1.0}, IVOA Recommendation to
  appear. \urlprefix\url{https://ivoa.net/documents/Coords/}

\bibitem[{{Salgado} et~al.(2013){Salgado}, {Osuna}, {Rodrigo}, {Allen},
  {Louys}, {McDowell}, {Baines}, {Maiz Apellaniz}, {Hatziminaoglou},
  {Derriere}, \& {Lemson}}]{2013ivoa.spec.1005S}
{Salgado}, J., {Osuna}, P., {Rodrigo}, C., {Allen}, M., {Louys}, M.,
  {McDowell}, J., {Baines}, D., {Maiz Apellaniz}, J., {Hatziminaoglou}, E.,
  {Derriere}, S., \& {Lemson}, G. 2013, {IVOA Photometry Data Model Version
  1.0}, IVOA Recommendation 05 October 2013. \eprint{1402.4752}

\end{thebibliography}
\end{document}